\documentclass[review,1p]{elsarticle}
\usepackage{bm}

\usepackage{amssymb}

\usepackage{color,ulem}

\journal{Physica E}

\begin{document}

\begin{frontmatter}



\title{Spintronics and spincaloritronics in topological insulators}


\author{Takehito Yokoyama and Shuichi Murakami}

\address{Department of Physics, Tokyo Institute of Technology, Tokyo 152-8551, Japan}

\begin{abstract}
We study spintronics and spincaloritronics in topological insulators.
We show spintronics effects in 2D topological insulator junctions and 3D topological insulators coupled to ferromagnets. We also investigate spin polarization on the surface of a topological insulator induced by a circularly polarized light and refraction at the junction between two topological insulators. 
As for spincaloritronics effects, we show transverse magnetic heat transport and thermoelectric transport in topological insulators. 
Finite-size effect in topological insulators and band structure engineering of interface states of topological insulators are also discussed.

\end{abstract}


\end{frontmatter}




\section{Introduction}
\label{introduction}
Recent discovery of topological insulator (TI) offers a new state of matter topologically different from the conventional band insulator.~\cite{Moore,Fu0,Schnyder,Qi2,Qi0,Hasan,Qi1} Edge channels or surface states of the TI are topologically protected, immune to small perturbations which respect time reversal symmetry,  and described by Dirac fermions at low energies. 
The surface Dirac fermion of TI has  been predicted to show interesting phenomena, e.g., superconducting proximity effect, in particular Majorana fermions\cite{Fu,Fu2,Akhmerov,Tanaka,Linder,Yokoyama6,Beenakker,Alicea}.
 
Since on the surface of the TI, electrons obey the (massless) Dirac equation, the spin and momentum of the electrons are coupled. Thus, one may expect novel spintronics and spincaloritronics effects  on the surface of the TI. \cite{Garate,Garate2,Mondal,Burkov,Nomura,Tserkovnyak,Mahfouzi,Mahfouzi2}

One salient feature of the Dirac fermions is that the Zeeman field acts like a vector potential. Therefore, one can expect anomalous spin related property by magnetic fields in TIs. We will see this feature in TIs attached to ferromagnets. The coupling between spin and momentum also gives novel coupling between 
spin transport and heat transport. This novel spin-heat coupling emerges not only as spincaloritronic effects, but also as thermoelectric effects. Namely, the 
coupling between charge transport and heat transport is also unconventional in TIs.

In this paper, we study spintronics and spincaloritronics in topological insulators.
We show spintronics effects in 2D topological insulator junctions and 3D topological insulators coupled to ferromagnets. We also investigate spin polarization on the surface of a topological insulator induced by a circularly polarized light and refraction at the junction between two topological insulators. 
As for spincaloritronics effects, we show transverse magnetic heat transport and thermoelectric transport in topological insulators. 
Finite-size effect in topological insulators and band structure engineering of interface states of topological insulators are also discussed.

\section{Spintronics}

In this section, we first consider spintronics effects in 2D TIs (or quantum spin Hall (QSH) systems ). Then, we study spintronics effects in 3D TIs. 
Transport properties of TIs have been nicely reviewed in Ref.\cite{Culcer}. 

\subsection{Spin rotation in normal metal/quantum spin Hall junctions}

\begin{figure}[htb]
\begin{center}
\scalebox{0.8}{
\includegraphics[width=9cm,clip]{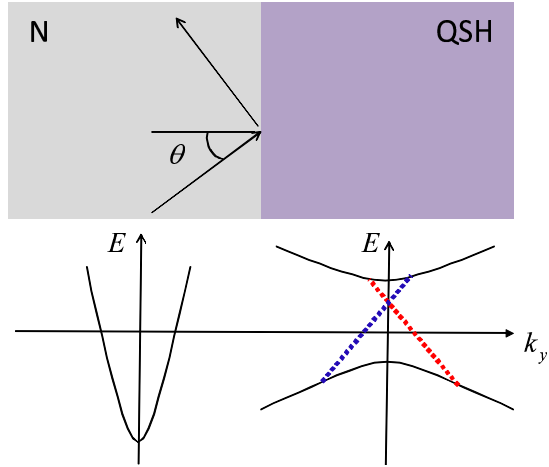}
}
\end{center}
\caption{N/QSH junction and corresponding band structures (below). Dotted lines represent helical edge modes. 
}
\label{fig2a}
\end{figure}

Let us commence with the scattering problem in normal metal/quantum spin Hall (N/QSH) junctions and show how the spin of an electron rotates upon scattering at the interface (see  Fig. \ref{fig2a}). The effective 4-band model proposed for HgTe/CdTe quantum wells is given by \cite{Konig2} 
\begin{eqnarray}
{\cal{H}}&=&\left(\begin{array}{cc} h(k)& 0\\
0&h^{*}(-k)\end{array}\right)\label{contH},
\end{eqnarray}
with $
h(k)=\epsilon (k) {\rm{I}}_{2\times 2}+d_a(k)\sigma^a,
\epsilon (k)=C-D(k_{x}^2+k_{y}^2), 
d_a (k)=\left(A k_x,-Ak_y,M(k)\right),
$ and
$M(k)= M-B(k_{x}^2+k_{y}^2)$ where we have used the basis order $({\vert
E_1 +\rangle,\vert H_1 +\rangle,\vert E_1 -\rangle,\vert H_1
-\rangle})$ ($``E"$ and $``H"$ represent the electron and hole bands, respectively), and,
$A,B,C,D,$ and $M$ are material parameters that depend on the quantum
well geometry.
 $\epsilon (k)$ is the averaged energy dispersion, while $M(k)$ is the energy difference between the valence and conduction bands. $A$ represents the strength of the band mixing due to the spin-orbit interaction. We note that 
${\cal{H}}$ is equivalent to 
two copies of the massive Dirac Hamiltonian but with a $k$-dependent mass $M(k)$.\cite{Konig2}  
In this system, the transition of electronic band structure occurs from a normal to an 
inverted type when the thickness of the quantum well is varied through a critical thickness. 
This corresponds to the sign change of the mass $M$ of this system.\cite{Bernevig2,Zhou} 
It should be noted that since $h(k) \neq h^{*}(-k)$ (which corresponds to the fact that chirality is different for different spins), spin orbit coupling has $z$ component, which results in spin rotation effect.

\begin{figure}[htb]
\begin{center}
\scalebox{0.8}{
\includegraphics[width=17.0cm,clip]{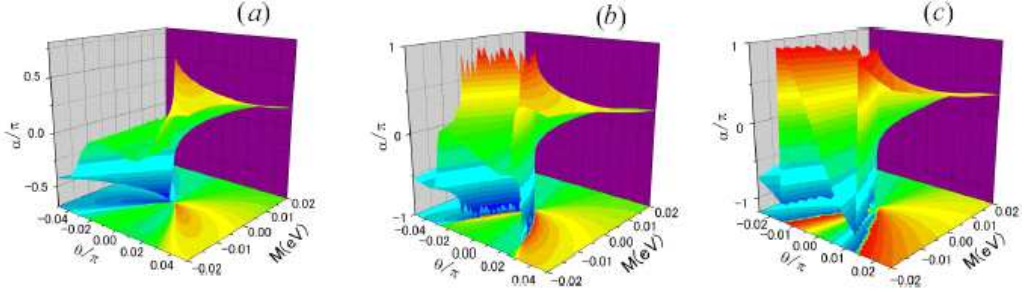}
}
\end{center}
\caption{Spin rotation angle $\alpha$ as a function of injection angle and $M$. 
(a) $C=-0.08$ eV, (b) $C=-0.1$ eV and (c) $C=-1$ eV. Note that $\alpha=\pi$ and $\alpha=-\pi$ are equivalent and the continuous increase of $\alpha$ in the clockwise direction is separated into several sheets in (b) and (c). Note that  $\alpha(-\theta)=-\alpha(\theta)$ is satisfied. 
}
\label{fig2b}
\end{figure}

By solving scattering problem in this junction with the use of the Hamiltonian Eq.(\ref{contH}), one can calculate the rotation angle of the spin of the electron within the plane $\alpha$. \cite{Yokoyama1}
Figure \ref{fig2b} shows the results for $\alpha$ in the plane of $(\theta,M)$ for $C=-0.08$ eV in (a), $C=-0.1$ eV in (b), and $C=-1$ eV in (c) with the other 
parameters fixed as $A=4$ $\rm eV \cdot $\AA, $B=-70$ $\rm eV \cdot $\AA$^2$ and  $D=-50$ $\rm eV\cdot $\AA$^2$. Here, $\theta$ is the angle of incidence as shown in
 Fig. \ref{fig2a}. These parameters are appropriate for the HgTe/CdTe quantum well. \cite{Konig2} 
In Figure \ref{fig2b} (a), a sharp ridge in $M<0$ and $\theta>0$ region 
and its negative correspondence in $M<0$ and $\theta<0$ are seen.
This is in stark contrast to the case of usual insulators $M>0$ 
although $\alpha$ is still nonzero there.  Note that the
height of the ridge is as high as $\sim \pi/2$.
With the increase of $|C|$, we find a qualitatively different structure. 
Near the origin in Fig. \ref{fig2b} (b), $\alpha$ reaches $\pi$, changes its sign, and \textit{winds by $4 \pi$ around the origin}, while it does not in the region far away from the origin. 
The angle at which the dispersions of the helical edge modes cross the  Fermi energy is given by 
$\theta _C  = \pm \sin ^{ - 1} \left[ {\frac{{MD}}{{Ak_F \sqrt {B^2  - D^2 } }}} \right].$\cite{Zhou}
We find that a large magnitude of $\alpha$ in Fig. \ref{fig2b} appears around this angle, which means that the helical edge modes resonantly enhance the spin rotation. 
A similar problem in 3D TIs has been also studied in Ref. \cite{Zhao}.

\subsection{Magnetoresistance in a ferromagnet/ferromagnet junction on the surface of a topological insulator}
\begin{figure}[htb]
\begin{center}
\scalebox{0.8}{
\includegraphics[width=15.0cm,clip]{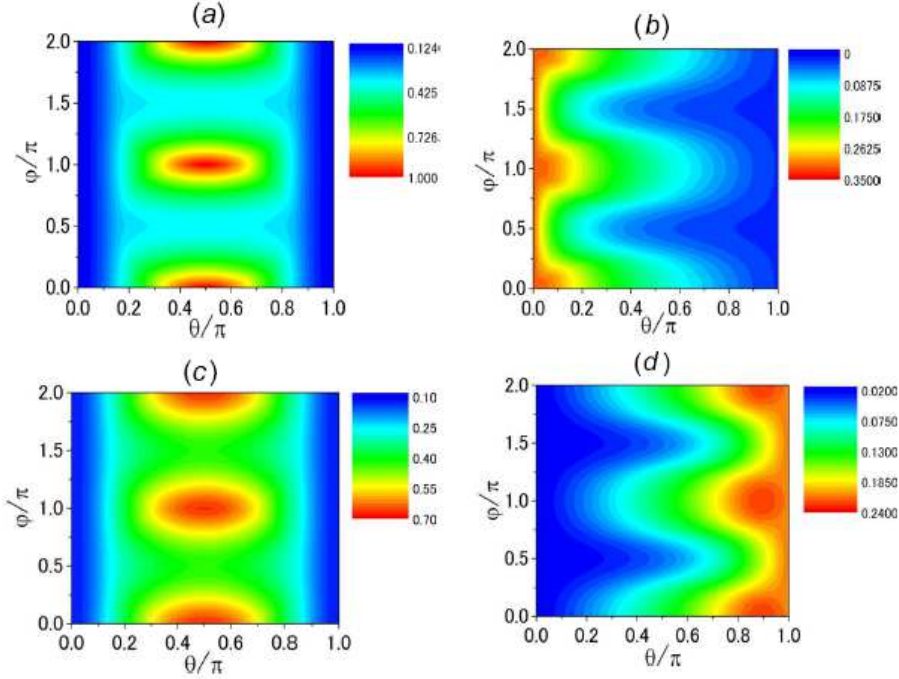}
}
\end{center}
\caption{Tunneling conductance $\sigma$ for $m_2=0$ ((a) and (c)), and  $m_2=\sqrt{0.9}E$ ((b) and(d)). n-n junction in (a) and (b). p-n junction in (c) and (d). 
}
\label{fig2c}
\end{figure}

Next, let us proceed to 3D TIs coupled to ferromagnets. 
We consider 2D ferromagnet/ferromagnet junctions which is abbreviated as F1/F2 in the following. We focus on charge transport at the Fermi level inside the bulk gap of the TI, which is described by the 2D massive Dirac Hamiltonian
\begin{eqnarray}
H = \left( {\begin{array}{*{20}c}
   {m_z } & {k_x  + m_x  - i(k_y  + m_y )}  \\
   {k_x  + m_x  + i(k_y  + m_y )} & { - m_z }  \\
 \end{array} } \right)
\end{eqnarray}
where $m_x, m_y$ and $m_z$ are exchange field and we set $v_F=\hbar=1$. Here, $x$-axis points to the interface normal while $y$-axis is parallel to the interface. We choose the exchange field in the F1 side as ${\bf{m}}_1$ $=(m_x ,m_y ,m_z ) = m_1 (\sin \theta \cos \varphi, \sin \theta \sin \varphi, \cos \theta)$ 
while in the F2 side, we set $m_x=m_y=0$ and $m_z=m_2$. 

By using the Landauer formula, one can calculate the tunneling conductance of the junction. \cite{Yokoyama2} Below, ``n" and ``p" mean that the Fermi level crosses the upper and the lower bands, respectively

Figure \ref{fig2c} depicts the normalized tunneling conductance $\sigma$ in n-n junction for (a) $m_2=0$ and (b) $m_2=\sqrt{0.9}E$ where $E$ is the Fermi energy. In Fig. \ref{fig2c} (a), the F2 is no longer ferromagnetic. However, the conductance strongly depends on the direction of the magnetization in the F1. 
At $\theta=0$ or $\pi$, the mismatch of the wavefunctions between the two sides  and that of the sizes of the Fermi surfaces suppress the conductance, because the energy $E$ is near the bottom of the upper band in F1 while there is no gap in F2. 
At $\theta=\pi/2$, on the other hand, the wavefunctions and the sizes of the Fermi surfaces are the same on both sides except the shift of Fermi surface in the momentum space due to the in-plane component of the magnetization. However, this misfit of the in-plane momentum between the two sides leads to a strong dependence of $\sigma$ on the in-plane rotation angle $\varphi$, which is not seen in the conventional magnetoresistance effect.
Since $k_y$ is conserved due to the translational symmetry along $y$-axis, 
the positions of the Fermi surfaces strongly affect the charge transport: 
if exchange field points to $x$-axis, there is no evanescent wave. 
On the other hand, when exchange field is applied in $y$-direction, the Fermi 
surface moves to the $k_y$ direction and hence the overlap region 
between $k_y$'s in the F1 and F2 is reduced. Therefore, the number 
of the evanescent modes increases and hence the conductance is strongly suppressed. 
Thus, we can obtain \textit{giant} magnetoresistance in this system. 
In Fig. \ref{fig2c} (b), the conductance is large at the parallel configuration ($\theta=0$) while it is small for antiparallel configuration ($\theta=\pi$) . 
Tunneling conductance in p-n junction is shown for $m_2=0$ in Fig.\ref{fig2c} (c) and  $m_2=\sqrt{0.9}E$ in Fig. \ref{fig2c} (d).
In Fig. \ref{fig2c} (d), in stark contrast to the conventional magnetoresistance effect, the conductance takes minimum at the parallel configuration ($\theta=0$) while it takes maximum  near antiparallel configuration ($\theta=\pi$).
These $\theta$ dependences can be understood by the overlap intergral of the 
wavefunctions on both sides.\cite{Yokoyama2}

\begin{figure}[htb]
\begin{center}
\scalebox{0.8}{
\includegraphics[width=7.0cm,clip]{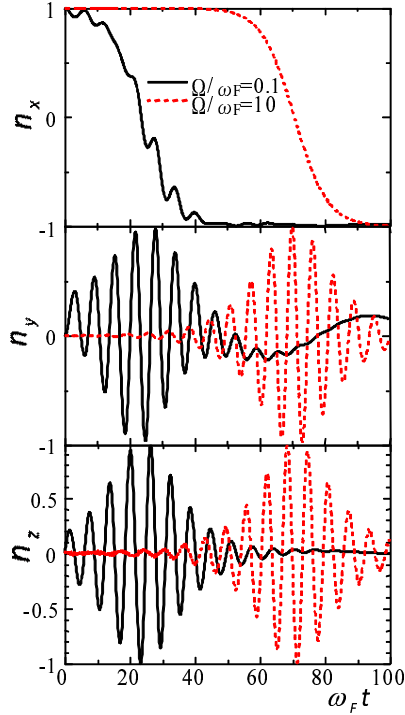}
}
\end{center}
\caption{Time evolution of the magnetization vector. $n_{x,y,z}$ is 3 components of the magnetization unit vector. $\Omega$ is the frequency of the applied voltage. 
}
\label{fig2d}
\end{figure}

\subsection{Magnetization dynamics on the surface of a topological insulator}

Up to here, we have considered TIs with static magnetizations. We next consider dynamical magnetizations.  
In Ref.\cite{Yokoyama3},  the dynamics of magnetization coupled to the surface Dirac fermions of a 3D TI has been investigated by deriving the Landau-Lifshitz-Gilbert equation in the presence of charge current. It is shown that both the inverse spin-galvanic effect and the Gilbert damping coefficient are related to the two-dimensional diagonal conductivity of the Dirac fermion, while the Berry phase of the ferromagnetic moment is related to the Hall conductivity. 
The derived Landau-Lifshitz-Gilbert equation has complicated torque terms, reflecting the spin structure of the surface Dirac fermion. 
Anomalous behaviors in various phenomena, e.g., the ferromagnetic resonance, are predicted in terms of this Landau-Lifshitz-Gilbert equation in Ref.\cite{Yokoyama3}.

Since there is a one-to-one correspondence between the direction of momentum and that of the spin on the surface of the TI,  an injected charge current can induce the magnetization, which is the so-called inverse spin-galvanic effect.\cite{Garate,Yokoyama3} The emergence of the inverse spin-galvanic effect is a direct consequence of the fact that, on the surface of TI, the velocity operator is given by the Pauli matrices in spin space. The current-induced magnetization on the surface exerts torque on the magnetization of the ferromagnet when they are noncollinear. 
Therefore, by injecting an electric current on the surface of the TI, one obtains spin transfer torque between the surface Dirac fermion and the magnetization. 

In Ref.\cite{Yokoyama5},  dynamics of the magnetization coupled to the surface Dirac fermions of a 3D TI has been studied by numerically solving the Landau-Lifshitz-Gilbert equation in the presence of charge current. The current-induced magnetization dynamics and the possibility of magnetization reversal have been discussed. Figure \ref{fig2d} shows an example of current-induced magnetization reversal from $n_x=1$ to $n_x=-1$. 
Since the magnetization dynamics is a direct manifestation of the inverse spin-galvanic effect, another ferromagnet is unnecessary to induce spin transfer torque in contrast to the conventional setup composed of two ferromagnets.\cite{Manchon}

\begin{figure}[htb]
\begin{center}
\scalebox{0.8}{
\includegraphics[width=10.0cm,clip]{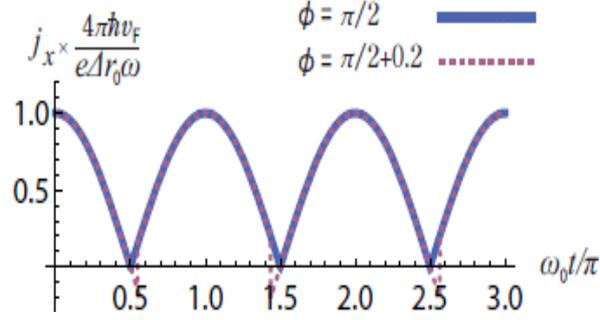}
}
\end{center}
\caption{The current on the surface of the TI as a function of time $t$ induced by the rotating exchange field.  $\phi$ is the angle of the precession axis measured from the normal to the interface. 
}
\label{fig2e}
\end{figure}

In Ref. \cite{Ueda}, 
a current dynamics on the surface state of a 3D TI induced by an exchange field of an attached ferromagnet has been discussed. It is found that the magnetic precession generates a direct charge current when the precession axis is within the surface plane as shown in Fig. \ref{fig2e}. This rectification effect is due to a quantum anomaly and is topologically protected. 
Intuitively, this phenomenon may be explained by an analogy to the quantum Hall effect: The exchange field perpendicular to the surface of TI plays a role of a magnetic field in the quantum Hall effect while that parallel to the surface corresponds to an applied inplane electric field. When the sign of the perpendicular exchange field changes, then that of the velocity of the parallel exchange field also changes. Therefore, simultaneous sign change of the magnetic field and the applied electric field does  not change the sign of the Hall current: the rectification effect. 
In Ref. \cite{Dora},  a similar rectification effect has been studied on an edge state of two-dimensional TIs induced by a circularly polarized light.

\subsection{The inverse Faraday effect on the surface of a topological insulator}

Optical effects of a TI have been studied theoretically in Refs.~\cite{Tse,Maciejko, Hosur}.
Due to the inverse Faraday effect\cite{Pitaevskii,van der Ziel,Pershan,Landau,Ede98}, spin polarization can be induced on the surface of TI by a circularly polarized light. 
In Ref. \cite{Misawa}, 
 the spin polarization of the electrons on the surface of TIs under  a circularly polarized light has been investigated. 
It is predicted that a light illumination can induce the out-of-plane component of the spin polarization as a result of the inverse Faraday effect. The magnitude of the spin polarization is proportional to the square of the lifetime $\tau$ and $\langle \sigma_{z}\rangle \simeq 2 \times 10^{-10}$
\AA
$^{-2}$ for typical parameters. 

\subsection{Refraction at the junction between two topological insulators}
\begin{figure}[htbp]
 \begin{center}
  \includegraphics[width=120mm]{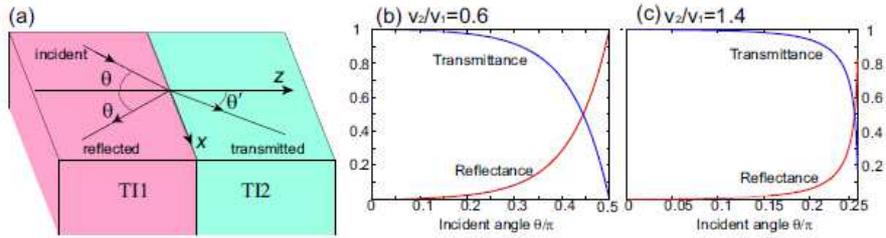}
 \caption{(a) Schematic of the refraction of the surface states
at the junction between the two TIs, TI1 and TI2. 
(b)(c): Reflectance (red) and transmittance (blue) for 
 the ratios of the velocities of the two TIs: (b)$v_{2}/v_{1}=0.6$ and (c)$v_{2}/v_{1}=1.4$.
}
\label{fig:kussetsu}
\end{center}\end{figure}

Because the surface states of TIs are spin-filtered, they are expected to 
show novel charge transport which cannot be realized in other systems. 
We note that the linear dispersion of the Dirac cone on the surface of the TI is similar to photons, and the velocity of the Dirac cone on the surface of 3D TI depends on materials: about $4\times10^5$m/s for Bi$_2$Te$_3$ \cite{Chen09}, and $5\times10^5$ m/s for Bi$_2$Se$_3$ \cite{Xia09}.  
Therefore, when two different TIs are attached, the refraction phenomenon similar to  
optics is expected at
the junction. The effective Dirac Hamiltonian of the surface states on the $xz$-plane is represented as $H=v({\bm \sigma} \times {\bf k})_y$, 
where $\sigma_i$ are the Pauli matrices, and $v$ is the Fermi velocity.
This Hamiltonian has the linear dispersion $E=\pm vk$ where $k=|{\bf k}|$ is the wavenumber. 
From this Hamiltonian we can easily see that for $v>0$ ($v<0$), the spin on the Fermi surface on the upper cone has the clockwise (counterclockwise) direction in ${\bf k}$ space.  Hence the sign of $v$ corresponds to the
spin chirality  in ${\bf k}$ space. 
In Ref.~\cite{TakahashiPRL}, a refraction problem is considered for a junction between the two TIs, which we call TI1 and TI2, with
 the incidence angle $\theta$, and the transmission
angle $\theta'$ (Fig.~\ref{fig:kussetsu}(a)).
Let $v_1$ and $v_2$ denote the velocities of the two TIs. 

We assume $v_1$ and $v_2$ to be positive. After a straightforward calculation, 
the transmittance
 and the reflectance are given as
\begin{equation}
R=\frac{\sin^2\frac{\theta'-\theta}{2}}{\cos^2\frac{\theta+\theta'}{2}},\ 
T
=\frac{\cos\theta\cos\theta'}{\cos^2\frac{\theta+\theta'}{2}}=1-R.
\end{equation}
The results are plotted in Fig.~\ref{fig:kussetsu}(b)(c). 
Unlike optics,  we have perfect transmission ($T=1$ and $R=0$) for normal incidence ($\theta=0$) \cite{TakahashiPRL}, which reflects the prohibited backscattering on the surface of the TI, corresponding to the Klein paradox proposed in graphene \cite{Katsnelson06}.

\section{Spincaloritronics}
Here, we focus on transverse and longitudinal heat transports in TIs in Secs. 3.1 and 3.2, respectively. 

\subsection{Transverse magnetic heat transport on the surface of a topological insulator}
In Ref. \cite{Yokoyama4}, 
 Nernst-Ettingshausen and thermal Hall effects have been investigated on the surface of a TI on which a ferromagnet is attached. General expressions of the Peltier and the thermal Hall conductivities are derived, which are reduced to  simple forms at low temperatures.  It is shown that the Peltier  and the thermal Hall conductivities show non-monotonous dependence on temperature. At low temperature, they have linear dependence on temperature. From the behavior of the Peltier conductivity  at low temperature, one can estimate the magnitude of the gap induced by time-reversal symmetry breaking. The Peltier conductivity can be used to map the Berry phase structure.

\subsection{Thermoelectric transport in topological insulators}
The edge states in 2D TIs have perfectly conducting channels like quantum Hall edge states. It is therefore expected to show anomalous thermoelectric transport as well. In Ref.~\cite{Takahashi} we proposed the 2D TI to be a potential candidate
to have efficient thermoelectric transport at low temperature.
Efficiency of thermoelectric energy conversions is characterized by
the thermoelectric figure of merit $ZT$. It is defined as $ZT= \sigma S^2T/\kappa$,
where $T$ is the temperature, $\sigma$ is the electrical conductivity, $S$ is the Seebeck coefficient, and $\kappa$ is the thermal conductivity. 
It is not easy to achieve large values of $ZT$, because 
$\sigma$, $S$ and $\kappa$ 
cannot be independently controlled.
We note here that some of the TIs such as 
Bi$_{1-x}$Sb$_x$ \cite{Hsieh},  Bi$_2$Se$_3$ \cite{Xia09}, and 
Bi$_2$Te$_3$ \cite{Chen09} are also  good thermoelectric materials. 

To calculate thermoelectric figure of merit, we introduce linear 
response of the electric current $j$ and thermal current $w$ to the electric field or thermal gradient in the following way: 
\begin{equation}
\left(\begin{array}{c}  
j/q \\ w \end{array}\right)=
\left(\begin{array}{cc} L_{0} & L_{1} \\ L_{1} & L_{2} \end{array}\right)
\left(\begin{array}{c}
 -\frac{\mathrm{d}\mu}{\mathrm{d}x} \\ -\frac{1}{T}\frac{\mathrm{d}T}{\mathrm{d}x} \end{array}\right),
\end{equation}
where $q$ is the charge $-e$, and $\mu$ is the chemical potential.
From the transport coefficients $L_j$, various 
thermal and electric properties are expressed as
\begin{eqnarray}
\sigma=e^2L_0,\ S = -\frac{1}{eT}\frac{L_1}{L_0},\
\kappa_e = \frac{1}{T}\frac{L_0 L_2 - L_1^2}{L_0},\
ZT = \frac{L_1^2}{L_0 L_2 - L_1^2 + \kappa_L T L_0},
\end{eqnarray}
where $\kappa_e$ and $\kappa_L$
are the thermal conductivity from electrons and phonons, respectively. We treat 
$\kappa_L$ as a constant.

\subsubsection{Two-dimensional topological insulators}
\begin{figure}[htbp]
\begin{center}
 \includegraphics[width=130mm]{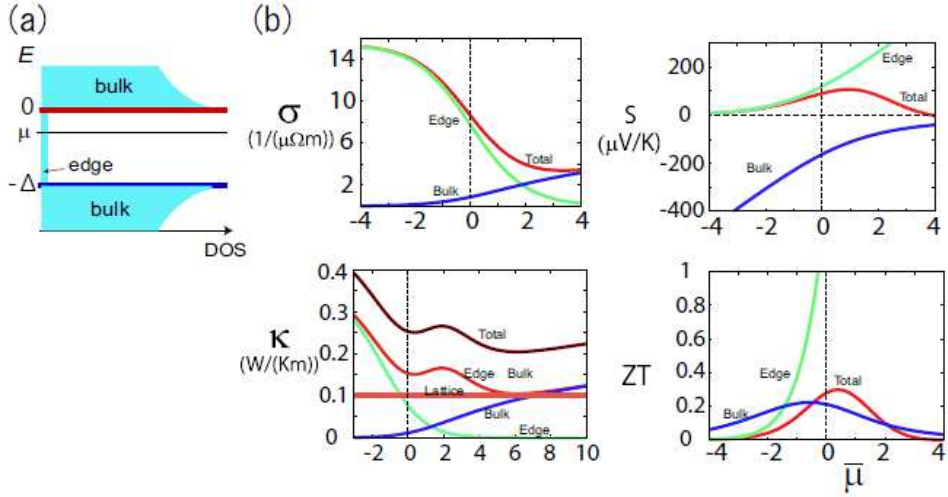}
 \end{center}
 \caption{(a) Schematic energy bands for the bulk and edge states used in the calculation. (b) shows the results. 
We employ physical quantities for Bi$_2$Te$_3$ as follows.
The electron effective mass is 0.02$m_e$ where $m_e$ is the free electron mass and the number of
 carrier pockets in the bulk is $c=6$.
The mobility $\mu^{*}$ is assumed to be 2000cm$^2$V$^{-1}$s$^{-1}$ at $T=$1.8K , and $\kappa_L$ is 0.1 Wm$^{-1}$K$^{-1}$. 
The width of the ribbon is 10nm.}
 \label{fig:WLproperties}
\end{figure}

In Ref.~\cite{Takahashi}, 
we showed that the edge states in a narrow ribbon of 2D TI give rise to large $ZT$ at low temperature. 
The edge channels of the 2D TIs 
are perfectly conducting, and it can be described by the Landauer formula  
with the transmission probability being unity. 
The density of states are schematically shown in
 Fig.~\ref{fig:WLproperties}(a), where $\Delta$ is the energy gap. 
We focus on energies around the edge of the bulk conduction band and
 neglect the valence band. 
In the calculation, bulk and edge transports are treated 
independently, which is valid within the inelastic scattering length $\ell$. Therefore, we can regard
the inelastic scattering length $\ell$ as an effective system size. 
The transport coefficients for the bulk are 
calculated by the Boltzmann equation.

When bulk and edge transport coexist, the total transport coefficients $L_\nu$ are given by their sum.
The Seebeck coefficients have opposite signs for the bulk and the edge transport, and that they cancel each other.  
As a result the bulk and edge thermoelectric transports compete each other, suppressing the $ZT$ in total \cite{Murakami10}.
The results are shown in Fig.~\ref{fig:WLproperties}. 
Maximum of $ZT$ occurs when $\mu$ is around 
the band edge because of the competition. 
We note that the edge transport coefficients are proportional to the inelastic scattering length $\ell$, while the bulk transport does not depend on $\ell$. 

At high temperature such as room temperature, $\ell$ is very short, and the bulk transport is dominant in the thermoelectric transport.
At low temperature, the edge transport becomes prominent because $\ell$ becomes long, and thermoelectric figure of merit is enhanced.
We note that for the perfect conduction for edge transport, 
the ribbon should be much wider than the penetration depth 
$\lambda$ of the edge states. 
The penetration depth $\lambda$ depends on the systems, and in some systems such as Bi ultrathin film, it is as short as the lattice constant \cite{Wada}.

In 3D TI systems, 
topologically protected 1D states   
are predicted on dislocations of the crystal, depending on the $Z_2$ topological numbers \cite{Ran}. 
These 1D states are perfectly 
conducting, and in Ref.~\cite{Tretiakov} we proposed that they lead to enhanced thermoelectric figure of merit at low temperature, especially for 
densely distributed dislocations in the crystal. The mechanism is 
similar to that for the edge states of 2D TIs.

\subsubsection{Three-dimensional topological insulators}
\begin{figure}[htc]
 \begin{center}
  \includegraphics[width=90mm]{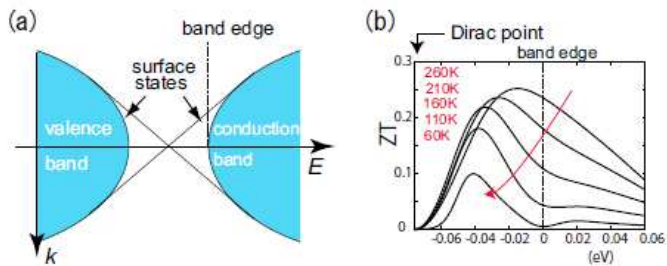}
 \end{center}
 \caption{(a) A schematic figure of the 3D TI band structure with gapless surface states, and (b) $ZT$ as a function of the chemical potential.}
 \label{fig:Figures3DTIgap}
\end{figure}

We discuss here thermoelectric properties of topological surface states 
in 3D TIs \cite{Takahashi2}.
However,
the surface states in 3D TIs are diffusive, unlike the edge states in 2D TI.
That reduces the surface transport, and 
leads to low thermoelectric figure of merit $ZT$ due to the competition with the bulk transport. 
We consider a thin slab 3D TI, and calculate transport coefficients along the slab 
for the surface states and the bulk states separately by the Boltzmann equation.
The phonon transport is neglected.
An example of the result for $ZT$ is shown in Fig~\ref{fig:Figures3DTIgap} (b). 
The result shows a competition between the surface and bulk states in the thermoelectric transport.
The surface transport dominates thermoelectricity at low temperatures, and
as the temperature is increased, the maximum value of the $ZT$ approaches the band edge. 

\section{Discussion}
\subsection{Finite-size effect in topological insulators}
\begin{figure}[htb]
\begin{center}
\scalebox{0.8}{
\includegraphics[width=10.0cm,clip]{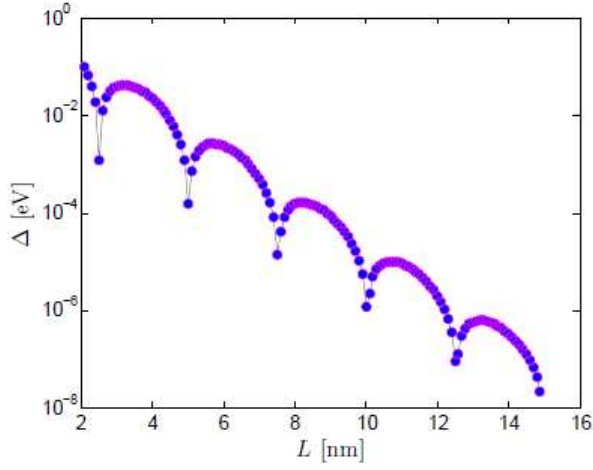}
}
\end{center}
\caption{Plot of the finite-size induced gap $\Delta$ between the surface states for Bi$_2$Se$_3$ as a function of the width $L$. 
}
\label{fig4a}
\end{figure}

Most of the intriguing effects discussed above stem from the (massless) Dirac fermion on the surface of TIs. However, in general, bulk electrons also contribute to the physics, which may mask the predicted effects. Therefore, to reduce the bulk contribution in transport properties is one of most important experimental tasks\cite{Culcer}.  To fabricate thin films of TIs seems most promising route to overcome this difficulty. However, one should keep in mind that when two surface states are sufficiently close to each other, namely within the decay length of the surface states, these states are overlapped and a gap opens in the surface states.\cite{Linder2,Liu2,Lu}

In Ref.\cite{Linder2}, it has been clarified
how the surface states in the strong TI Bi$_2$Se$_3$ are influenced by finite size effects using the effective Hamiltonian for Bi$_2$Se$_3$. It is found that the surface-states in Bi$_2$Se$_3$ display a remarkable robustness towards decreasing the width  $L$ down to a few nm, thus ensuring that the topological surface states remain intact, and that the gapping due to the hybridization of the surface states features  an oscillating exponential decay as a function of $L$ in Bi$_2$Se$_3$ as shown in Fig. \ref{fig4a}. 

To reduce the bulk contribution, gate voltage may be useful.\cite{Chen} 
In addition to carrier tuning, gate voltage may form 1D states.  In Ref. \cite{Yokoyama7}, the formation of one-dimensional channels on the topological surface under the gate electrode has been predicted. This can be understood by regarding a momentum in the perpendicular direction as a ``mass". The energy dispersion of these channels is almost linear in the momentum, and its velocity and sign sensitively depend on the strength of the gate voltage. As a result, the local density of states near the gated region has an asymmetric structure with respect to zero energy.
In the presence of the electron-electron interaction, the correlation effect can be tuned by the gate voltage. 

\subsection{Band structure engineering of interface states of topological insulators}
\begin{figure}[htbp]
\begin{center}
\includegraphics[width=60mm,clip]{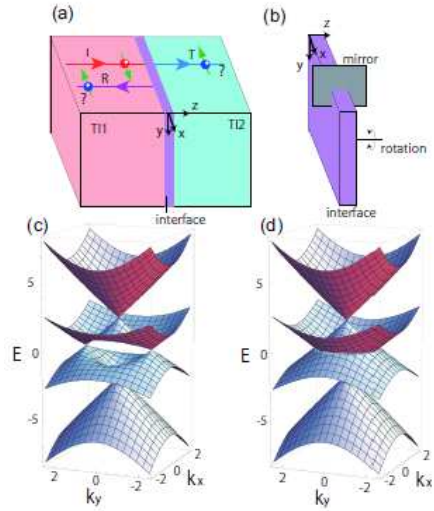}
\end{center}
\caption{(a) Schematic figure of refraction on the junction between the two TIs (TI1 (red) and TI2 (blue)),  whose velocities have different signs. The purple region represents the interface between the two TIs. (b) For the interface (purple), we consider mirror symmetry with respect to the $yz$ plane, and some kind of rotations around the $z$ axis. (c)(d) Dispersion on the interface between the two TIs with different signs of velocities.
 In (c), no rotational symmetry is imposed while in (d) the continuous rotational symmetry around the $z$ axis is imposed. }
\label{fig:TIjunction}
\end{figure}

So far we have seen various novel properties of spin transport in 
topological insulator surfaces. The surface band structure typically has 
a single Dirac cone, with its velocity determined by the material itself. 
The band structure can be designed further, by making interfaces between 
two TIs, as we see in the following \cite{Takahashi}. 
We reconsider the refraction problem at the junction between two TIs in Section 2.
When the velocities of the two TIs have opposite signs.
Then both reflection and transmission for the surface transport is prohibited for normal incidence,  
because
they violate spin conservation (see Fig.~\ref{fig:TIjunction}(a)).
Our answer to this paradox is the 
following. 
Here we show existence of gapless states at the interface between the two TIs
(the purple region in Fig.~\ref{fig:TIjunction}(a)) \cite{Takahashi}.
The normally incident wave goes along the surface of one TI, then flows into the interface.

These interface states arise from hybridization between
the two surface states from the two TIs.
To show the existence of gapless interface states, we first 
write down the effective Hamiltonian at 
the interface from the two Dirac cones with 
hybridization:
\begin{eqnarray}
H=\left(
\begin{array}{cc}
H_{1}&V\\
V^{\dagger}&H_{2}
\end{array}\right),\ \ H_{1}=v_{1}( {\bm{\sigma}} \times{\bf k})_z
, \ 
H_{2}=-v_{2}({\bm{\sigma}} \times{\bf k})_z
 \label{eq:mhamiltonian}
\end{eqnarray}
Here 
$H_{1(2)}$ is the effective surface Hamiltonian for the
surface of TI1 (TI2) at the interface,
and $V$
is the hybridization.
For simplicity, we retain only the lowest order in ${\bf k}$.
We henceforth impose the mirror symmetry with respect to the $yz$ plane, ${\cal M}_{yz}$, which restricts the form of the hybridization matrix $V$.
A direct calculation from the Hamiltonian (\ref{eq:mhamiltonian}),  we can see that for $v_1v_2>0$ the interface states 
are gapped by the hybridization, and for $v_1v_2<0$ 
there are gapless states on the interface. 
Dispersion of the gapless states depends on spatial symmetries:
\renewcommand{\labelenumi}{(\roman{enumi})}
\begin{enumerate}
\item No rotational symmetry: the interface states have two Dirac cones (Fig.~\ref{fig:TIjunction}
(d)).  
\item Continuous rotational symmetry: the gap of the interface states is closed 
along a loop in ${\bf k}$ space (Fig.~\ref{fig:TIjunction}
(c)).  
\item Discrete crystallographic rotational symmetry: the interface states have a collection of Dirac cones, whose number depends on details of the rotational symmetry. 
For $C_3$ symmetry there are six Dirac cones.
\end{enumerate}

The existence of gapless interface states can also be shown in the following way. 
Each TI with mirror symmetry ${\cal M}_{yz}$ is characterized by the mirror Chern number \cite{Teo08}.
In TI1 and TI2, the two surface Dirac cones have opposite velocities, and it 
corresponds to different mirror Chern numbers;
the TI1 has $n_{\mathcal{M}}^{(1)}=-1$, and 
the TI2 has $n_{\mathcal{M}}^{(2)}=1$. 
This difference by two gives rise to the two Dirac cones on the mirror plane for the interface states.
These
interface states do not come from the Z$_2$ topological number, but come from the mirror Chern number, 
and are protected by the mirror symmetry ${\cal M}_{yz}$.
Therefore it is a new class 
of the surface states originated from the mirror symmetry, similar to the topological 
crystalline insulators \cite{Fu11,Hsieh12}.
Both in the present work of interface states and in the topological crystalline insulators \cite{Fu11,Hsieh12}, the interface/surface states have even number of Dirac 
cones, and they are protected by mirror symmetry. In the present work Dirac cones appear on an interface between two $Z_2$-nontrivial TIs, while in the topological crystalline insulator they appear on a surface of a $Z_2$-trivial insulator.

We note that in a related work 
\cite{Apalkov} on an interface between two TIs with the same signs of 
velocities, a surface-state dispersion is reported to have an infinite slope (velocity), which is called tachyons. We think that such an infinite slope cannot occur in TI interfaces, since the velocity given by 
${\bf{v}}=\frac{dE}{d{\bf k}}=\left\langle \Psi \left| \hat{{\bf v}}\right|\Psi\right\rangle
$ ($|\Psi\rangle$: eigenstate) cannot diverge even for surface states. 
The whole eigenvalue problem (i.e. the Schr\"odinger equation plus decaying 
boundary conditions at $z\rightarrow \infty$) is definitely Hermitian. This
hermiticity forbids appearance of such tachyon-like modes in the TI junction problems.

\section{Summary}

In this paper, we have investigated spintronics and spincaloritronics in topological insulators.
We showed spintronics effects in 2D topological insulator junctions and 3D topological insulators coupled to ferromagnets. We also investigated spin polarization on the surface of a topological insulator induced by a circularly polarized light and refraction at the junction between two topological insulators. 
As for spincaloritronics effects, we showed transverse magnetic heat transport and thermoelectric transport in topological insulators. 
Finite-size effect in topological insulators and band structure engineering of interface states of topological insulators are also discussed.

\section{Acknowledgments}
We thank Y. Tanaka, N. Nagaosa, J. Zang, R. Takahashi, H. T. Ueda, A. Takeuchi, G. Tatara, T. Misawa, A. V. Balatsky, J. Linder, and A. Sudb{\o} for fruitful collaborations.

This work is partly supported by Grant-in-Aids from Asahi Glass foundation, MEXT, Japan (No. 21000004 and No. 22540327), and by the Global Center of Excellence Program by MEXT, Japan, through the ``Nanoscience and Quantum Physics" Project of the Tokyo Institute of Technology and also by a Grant-in-Aid for Young Scientists (B) (No. 23740236) and the ``Topological Quantum Phenomena" (No. 23103505) Grant-in-Aid for Scientific Research on Innovative Areas from theMinistry of Education, Culture, Sports, Science, and Technology (MEXT) of Japan.











\end{document}